\documentclass[%
 reprint,
 superscriptaddress,
 amsmath,amssymb,
 aps,
 pra,
]{revtex4-2}

\usepackage{graphicx}
\usepackage{dcolumn}
\usepackage{bm}
\usepackage{braket} 
\usepackage{siunitx}

\begin{document}

\preprint{APS/123-QED}

\title{Mode-Selective Micromotion Compensation via Coherent Excitation of a Ground-State-Cooled Motional Mode}

\author{Tetsu Kinase}
\email{tetsukinase@gmail.com}
\affiliation{
Komaba Institute for Science (KIS), The University of Tokyo, Meguro-ku, Tokyo, 153-8902, Japan
}

\author{Ippei Nakamura}
\affiliation{
Komaba Institute for Science (KIS), The University of Tokyo, Meguro-ku, Tokyo, 153-8902, Japan
}

\author{Atsushi Noguchi}%
\email{u-atsushi@g.ecc.u-tokyo.ac.jp}
\affiliation{
Komaba Institute for Science (KIS), The University of Tokyo, Meguro-ku, Tokyo, 153-8902, Japan
}
\affiliation{
RIKEN Center for Quantum Computing (RQC), Wako, Saitama 351–0198, Japan
}
\affiliation{Inamori Research Institute for Science (InaRIS), Kyoto-shi, Kyoto 600-8411, Japan}

\date{\today}

\begin{abstract}

Excess micromotion induced by stray DC electric fields limits the control fidelity of trapped-ion experiments. Various techniques have been developed to detect and compensate ion displacements from the RF null by monitoring modulated fluorescence, ion positions, or micromotion sidebands. Although these approaches are effective, they do not probe the displacement through a ground-state-cooled motional mode and therefore offer limited compatibility with quantum-enhanced sensing protocols. Here, we demonstrate a micromotion compensation method based on the driven coherent amplitude of a ground-state-cooled motional mode. In this approach, the ion displacement is converted into a coherent motional excitation by an auxiliary RF drive and detected through the qubit. We determine the optimal compensation field with a $1\sigma$ statistical uncertainty of $\SI{8}{mV/m}$. The method is naturally compatible with quantum-enhanced sensing protocols employing nonclassical motional states, including Fock, squeezed, and Schrödinger-cat states, providing a pathway toward quantum-enhanced micromotion compensation.

\end{abstract}

\maketitle


\section{\label{sec:level1} Introduction}

Trapped ions provide a well-isolated platform for quantum science and technology \cite{bruzewicz2019trapped,blatt2012quantum,gill2003trapped}. Their long-lived internal states can be controlled with laser and microwave fields \cite{leibfried2003quantum,ballance2016high,quinn2026high,harty2014high,srinivas2021high,loschnauer2025scalable}, while laser cooling confines their motion to  well below an optical wavelength \cite{eschner2003laser,xing2025trapped,clements2026sub} and enables preparation near the motional ground state \cite{king1998cooling,schulz2008sideband,huang2024electromagnetically, wu2025electromagnetically}. These capabilities underpin optical atomic clocks \cite{burt2021demonstration,marshall2025high}, quantum sensing \cite{mccormick2019quantum, bonus2025ultrasensitive}, quantum simulation \cite{qiao2024tunable,guo2024site}, and quantum information processing \cite{moses2023race,ransford202698}.

In trapped ion quantum processors, ions are typically trapped in linear Paul traps by a combination of radio frequency (RF) and static electric fields \cite{raizen1992ionic}. A major challenge in Paul traps arises from stray DC electric fields. These fields displace the ion from the RF null and produce driven motion at the trap RF frequency. This motion, known as excess micromotion \cite{berkeland1998minimization}, periodically modulates the ion-laser interaction through the Doppler effect. For example, micromotion degrades Doppler-cooling performance and limits the minimum achievable temperature. In addition, previous studies have reported that the displacement from the RF null can also increase motional heating \cite{kalincev2021motional}, compromising the preparation and control of low-energy motional states. Accurate  micromotion compensation is therefore essential for trapped-ion experiments \cite{keller2015precise, pyka2014high}.

Several established techniques detect excess micromotion or the displacement that causes it. Photon-correlation measurements determine the micromotion amplitude from correlations between fluorescence and the trapping RF signal \cite{berkeland1998minimization, keller2015precise,pyka2014high}. On narrow optical transitions, micromotion can be measured through strengths of micromotion sidebands or changes in the carrier Rabi frequency induced by phase modulation \cite{goham2022resolved,lee2023micromotion, arnold2024enhanced}. Position-based methods identify the RF null by monitoring the ion's position while varying the trapping RF amplitude \cite{saito2021measurement, saito2025three}. Parametric excitation through the trap electrodes
provides another displacement-sensitive probe \cite{ibaraki2011detection, tanaka2012micromotion, nadlinger2021micromotion}.  In particular, modulation near a secular frequency can generate coherent motion whose amplitude depends linearly on the stray field, with fluorescence used for
detection \cite{nadlinger2021micromotion}. 
While these techniques provide effective means of micromotion compensation, they generally do not probe the displacement from the RF null through a ground-state-cooled motional mode that can be coherently controlled and read out via the ion qubit. In contrast to conventional parametric-excitation methods that detect driven motion through fluorescence \cite{nadlinger2021micromotion}, qubit-assisted sideband spectroscopy enables coherent motional excitations to be characterized through their phonon-number distribution. This provides a quantum-state-resolved approach to displacement sensing. Furthermore, the ability to prepare and measure motional quantum states offers a natural route toward quantum-enhanced sensing protocols employing Fock states \cite{wolf2019motional}, squeezed states \cite{burd2019quantum}, and Schrödinger-cat states \cite{hempel2013entanglement}, which have been shown to enhance sensitivity to small coherent displacements. Mapping the displacement from the RF null onto a controllable motional degree of freedom therefore provides a pathway toward quantum-enhanced micromotion detection.

Here, we develop a method that probes the ion displacement from the RF null through a ground-state-cooled motional mode of a trapped ion. We apply an auxiliary oscillating voltage to the RF electrodes at a frequency near the secular resonance, generating a displacement-dependent force that coherently excites the selected motional mode. By tuning the auxiliary drive frequency to a selected secular mode, the method provides information about the ion displacement along the corresponding motional direction. This directional information can facilitate the identification and compensation of stray electric-field components along different motional directions. The resulting coherent excitation is characterized through qubit-assisted sideband spectroscopy. Because the displacement is encoded in a coherently controllable motional state, the method also provides a natural starting point for incorporating nonclassical motional states into micromotion compensation. We demonstrate the method along one radial direction of a single $^{88}\rm{Sr}^{+}$ ion and determine the optimal compensation field with a statistical uncertainty of $\SI{8}{mV/m}$.

\section{\label{sec:level2} Theory}

\begin{figure}[b]
\includegraphics[keepaspectratio, height=0.4\textwidth]{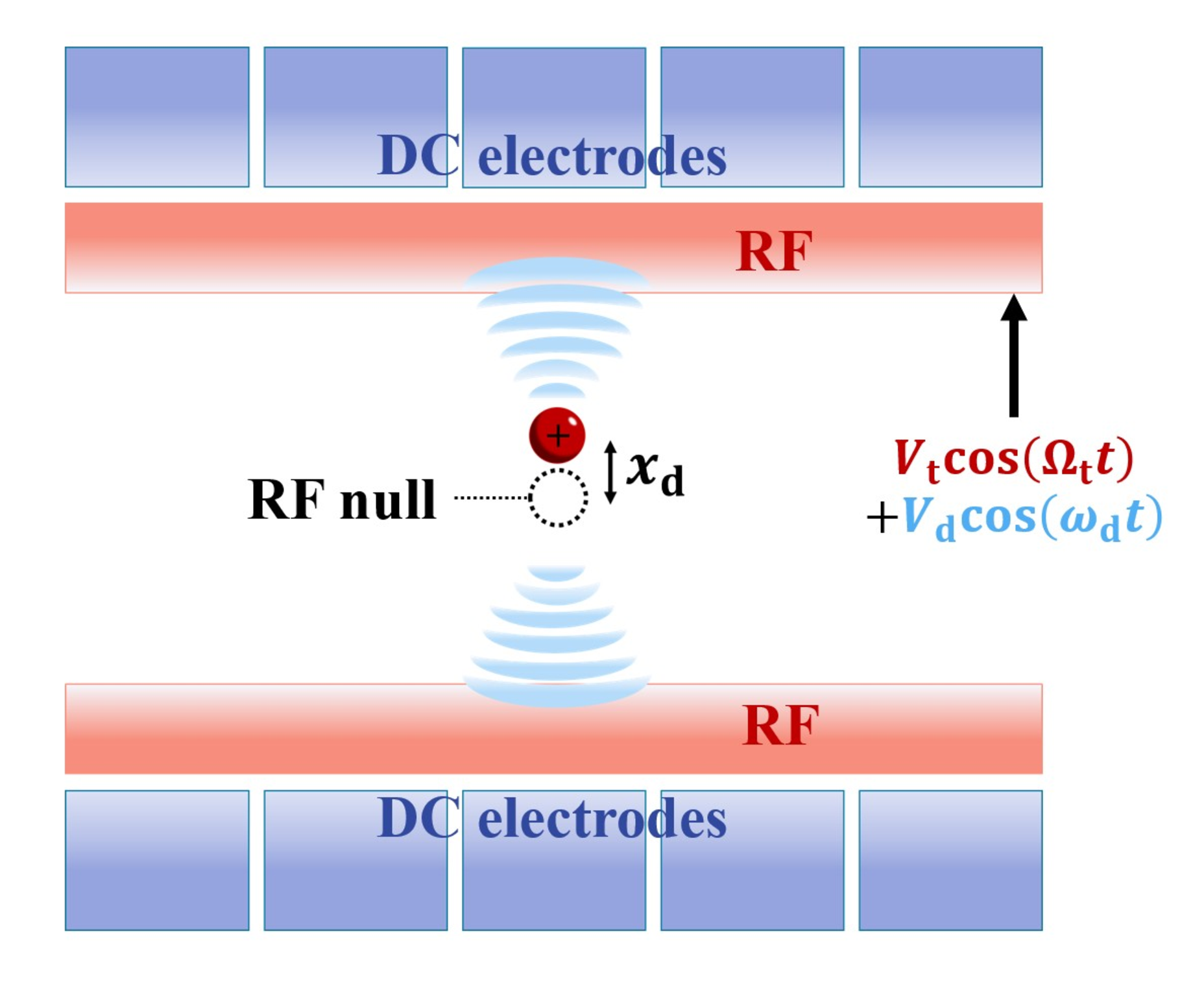}
\caption{\label{figure1}  Principle of the displacement-sensitive drive. Red and blue electrodes denote RF and DC electrodes, respectively. An auxiliary voltage applied to the RF electrodes generates a force proportional to the ion displacement from the field null and drives the selected secular mode.
}
\end{figure}

To detect the stray DC electric field, our method employs two RF voltages at different frequencies applied to the trap RF electrodes (Fig.~\ref{figure1}). One is the trapping RF drive with amplitude $V_{\rm{t}}$ and frequency $\Omega_{\rm{t}}$, while the other is an additional RF drive with amplitude $V_{\rm{d}}$ and frequency $\omega_{\rm{d}}$. We consider the regime $V_{\rm{t}} \gg V_{\rm{d}}$ and focus on a single radial coordinate $x$. The oscillating electric field is
\begin{equation}
\label{Equation1}
E(x,t) = 2G\lbrack V_{\rm{t}}\cos(\Omega_{\rm{t}}t)+V_{\rm{d}}\cos(\omega_{\rm{d}}t)\rbrack x. 
\end{equation}
Here, $G$ is the geometric factor of the electrode, with units of $\rm{m}^{-2}$. A stray DC field shifts the equilibrium position by $x_{\rm{d}}$. Writing $x = \delta x + x_{\rm{d}}$, where $\delta x$ is the displacement from the shifted equilibrium, gives
\begin{eqnarray}
\label{Equation2}
E(\delta x, t) =&& 2G\lbrack V_{\rm{t}}\cos(\Omega_{\rm{t}}t)+V_{\rm{d}}\cos(\omega_{\rm{d}}t)\rbrack(\delta x + x_{\rm{d}}) \nonumber\\
=&& 2G\lbrack V_{\rm{t}}\cos(\Omega_{\rm{t}}t)\delta x+V_{\rm{t}}\cos(\Omega_{\rm{t}}t)x_{\rm{d}}\nonumber\\
&&+V_{\rm{d}}\cos(\omega_{\rm{d}}t)\delta x+V_{\rm{d}}\cos(\omega_{\rm{d}}t)x_{\rm{d}}\rbrack.
\end{eqnarray}
The first term produces the trapping pseudopotential, and the second term drives excess micromotion induced by the ion displacement. The third term is an additional time-dependent field gradient and can modify the secular frequency through the two-frequency Paul-trap dynamics \cite{trypogeorgos2016cotrapping} discussed in Appendix G. The fourth term produces a force proportional to $x_{\rm{d}}$. When the drive frequency $\omega_{\rm{d}}$ is tuned close to the radial secular frequency $\omega_{x}$, this force resonantly excites the secular motion and coherently displaces the motional state in phase space. The resulting coherent-state amplitude $|\alpha|$ produced by a Gaussian pulse with temporal width $\sigma_{\rm{d}}$ is given by
\begin{equation}
\label{Equation3}
|\alpha(\Delta)|=\frac{\sqrt{2\pi}G|q|V_{\rm{d}}|x_{\rm{d}}|x_{\rm{zpf}}}{\hbar}\sigma_{\rm{d}}\rm{exp}(-\Delta^{2}\sigma_{\rm{d}}^{2}/2)
\end{equation}
where $q$ and $m$ are the ion charge and mass, $\Delta$ is the frequency detuning from the secular resonance, and $x_{\rm{zpf}}=\sqrt{\frac{\hbar}{2m\omega_{x}}}$ is the zero-point fluctuation. Equation~(\ref{Equation3}) shows that $|\alpha|$ is proportional to the displacement from the RF null. The optimal DC compensation field is therefore identified by minimizing $|\alpha|$.

We determine $|\alpha|$ from the blue-sideband Rabi oscillations using two models: an approximate sinusoidal fit at large displacement and a phonon-distribution model near the compensation minimum. For large coherent excitations, we first approximate the blue-sideband dynamics by a sinusoidal oscillation. In a displaced frame, the motional state annihilation operators can be replaced by $\hat{a} \rightarrow \hat{a}' +|\alpha|e^{-i\phi}$ where $\phi$ is the phase of the coherent displacement, and the blue-sideband interaction Hamiltonian is
\begin{eqnarray}
    \hat{H}_{\rm{I}} &&= \hbar\frac{\eta_{x}\Omega_{0}}{2}\lbrack\hat{\sigma}_{+}(\hat{a}'^{\dagger} +|\alpha|e^{i\phi})+\hat{\sigma}_{-}(\hat{a}' +|\alpha|e^{-i\phi})\rbrack \nonumber \\
    &&=\hbar\frac{\eta_{x}\Omega_{0}}{2}\big[(\hat{\sigma}_{+}\hat{a}'^{\dagger} +\hat{\sigma}_{-}\hat{a}') +|\alpha|(\hat{\sigma}_{+}e^{i\phi} +\hat{\sigma}_{-}e^{-i\phi})\big] \nonumber. 
\end{eqnarray}
Here, $\eta_{x}$ is the Lamb-Dicke parameter along the $x$ direction. For $|\alpha|\gg 1$, the terms proportional to $|\alpha|$ dominate, giving
\begin{equation}
    \label{Equation4}
    \hat{H}_{\rm{I}}
    \cong \hbar\frac{\eta_{x}\Omega_{0}}{2}|\alpha|(\hat{\sigma}_{+}e^{i\phi} +\hat{\sigma}_{-}e^{-i\phi}) .
\end{equation}
Thus, in the large-amplitude regime and within the Lamb-Dicke approximation, the effective blue-sideband Rabi frequency scales linearly with the coherent-state amplitude \(|\alpha|\). The ratio of the effective Rabi frequency after the drive to the \(n=0\) blue-sideband Rabi frequency therefore provides an approximate estimate of \(|\alpha|\). However, this linear scaling relies on the Lamb-Dicke approximation, under which the blue-sideband Rabi frequency for a Fock state \(|n\rangle\) is proportional to \(\sqrt{n+1}\). For sufficiently large coherent excitations, the Lamb-Dicke condition may fail, and higher-order terms in the ion-laser coupling can modify the sideband Rabi-frequency scaling. Furthermore, the sinusoidal approximation becomes insufficient near the compensation minimum, where the coherent-state amplitude is small. In this regime, another model is utilized to estimate $|\alpha|$. Assuming that the ion is prepared in a displaced thermal motional state, the blue-sideband excitation probability is
\begin{eqnarray}
    \label{Equation5}
    &&{P}_{\uparrow}(t) = \frac{p_{\rm{init}}}{2}\sum_{n=0}^{N}
    p_{n}(\alpha,n_{\rm{th}})\big\lbrack 1-e^{-\gamma t}\cos(\eta_{x}\Omega\sqrt{n+1}t) \big\rbrack\nonumber\\
    &&p_{n}(\alpha,n_{\rm{th}})=\bra{n}D(\alpha)\rho_{\rm{th}}(n_{\rm{th}})D^{\dagger}(\alpha)\ket{n}
\end{eqnarray}
where $p_{\rm{init}}$ is the spin-initialization probability, $n_{\rm{th}}$ is the initial mean phonon number, \(N\) is the phonon-number cutoff chosen to retain the relevant phonon-number states, and \(\gamma\) describes the damping of the Rabi oscillations. The parameters $p_{\rm{init}}$, $n_{\rm{th}}$ and $\eta_{x}\Omega$ are calibrated without an auxiliary-drive pulse (Appendix C). This model can be used to estimate \(|\alpha|\) over a wide range of coherent-state amplitudes, provided that the Lamb-Dicke approximation remains valid. Unlike the large-$|\alpha|$ approximation, this
model retains the phonon-number distribution and can describe small coherent excitations, subject to the validity of the sideband-coupling model.

\begin{figure}[b]
\includegraphics[keepaspectratio, height=0.7\textwidth]{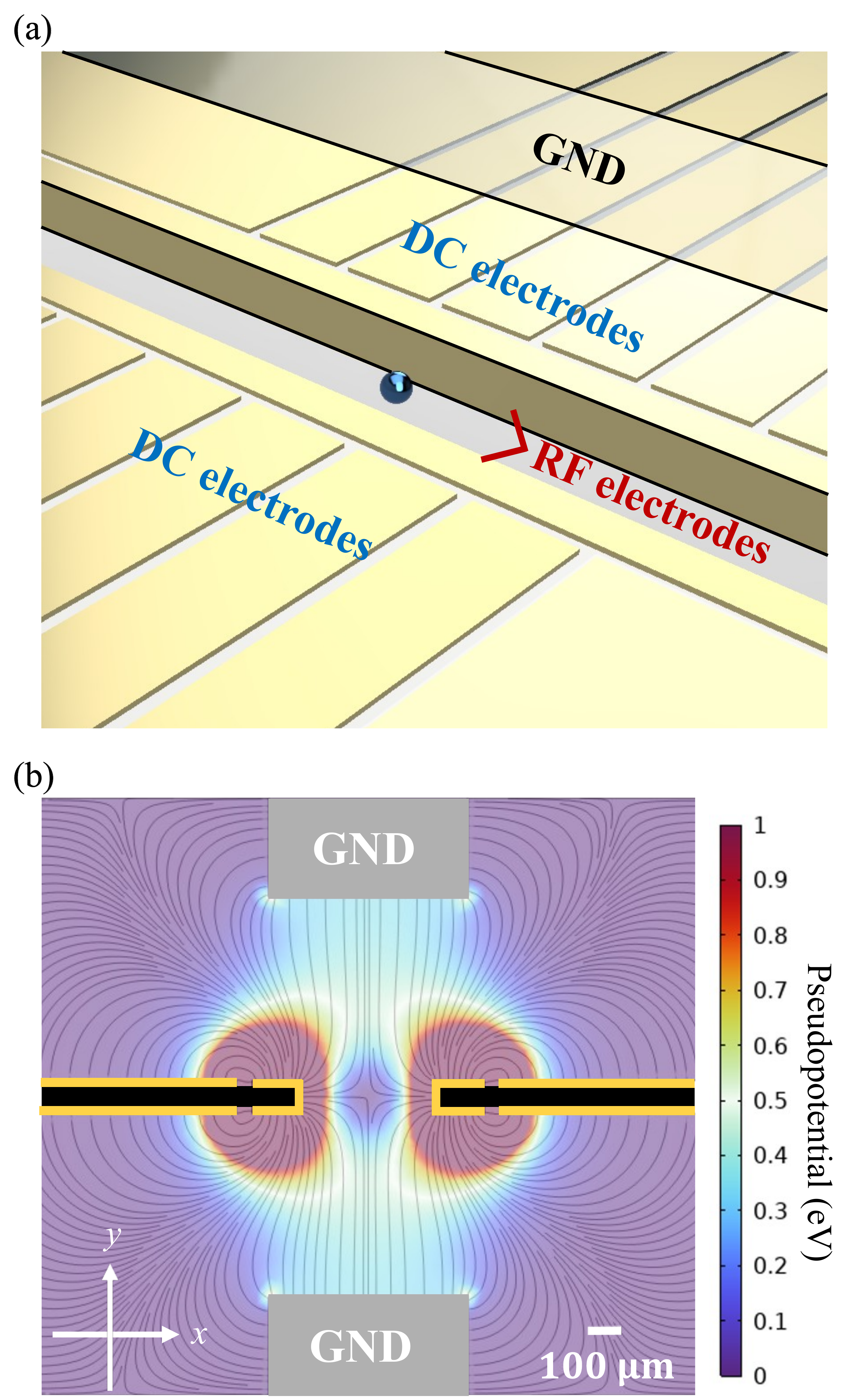}
\caption{\label{figure2} (a) Trap electrode geometry. The monolithic chip has a metallized slit forming the RF electrode surfaces, segmented DC electrodes, and ground electrodes above and below the trapping region. (b) Simulated trapping pseudopotential. The color scale gives the pseudopotential in eV, and the lines show the electric-field lines.
}
\end{figure}


\section{\label{sec:level3} Experimental results}
The experiments were performed in a room-temperature ultrahigh-vacuum chamber using a linear Paul trap. The trap features a monolithic electrode geometry similar to that reported in \cite{luo2026monolithic}, as shown in Fig.~\ref{figure2}(a). The trap chip was fabricated from a liquid crystal polymer (LCP) substrate with gold-plated electrodes. A  $\SI{10}{\milli m} \times \SI{0.4}{\milli m}$ slit was bored in the chip and metallized along its inner surface to form the RF electrode. Segmented DC electrodes  surrounded the slit. The ion was confined  near the center, at an ion-electrode distance of $d=\SI{200}{\micro m}$. Fabricating both RF electrode surfaces on a single substrate defined their relative alignment through the chip fabrication process.
Figure~\ref{figure2}(b) shows the pseudopotential numerically calculated by COMSOL Multiphysics\textsuperscript{\textregistered} \cite{COMSOL}. For an RF amplitude of $\SI{200}{V}$, the numerically calculated trap depth is approximately $\SI{0.4}{eV}$.

\begin{figure*}
\includegraphics[keepaspectratio, height=0.6\textwidth]{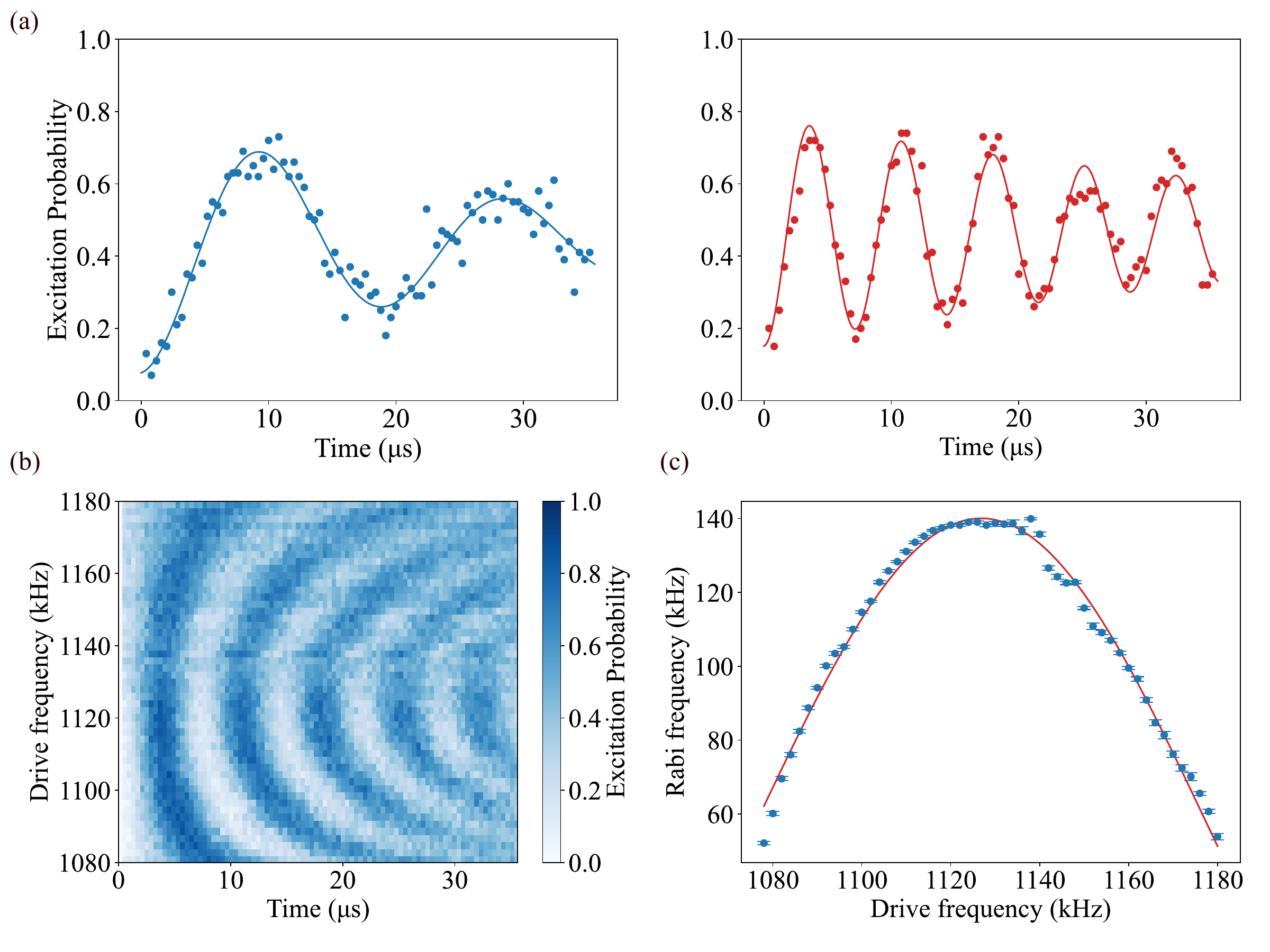}
\caption{\label{figure3} (a) Blue-sideband Rabi oscillations following an off-resonant drive at $\omega_{\rm{d}}=2\pi\times\SI{1080}{kHz}$ (blue circles) and a near-resonant drive at $\omega_{\rm{d}}=2\pi\times\SI{1126}{kHz}$ (red circles). Sinusoidal fits give $\eta_{x}\Omega_{0}|\alpha|  = 2\pi\times\SI{52}{kHz}$ and $2\pi\times\SI{139}{kHz}$, respectively. (b) Spin-excitation probability versus drive frequency and blue-sideband pulse duration. (c) Effective Rabi frequencies extracted from (b), with a fit to the Gaussian frequency response in Eq.~(\ref{Equation3}).
}
\end{figure*}

Neutral strontium atoms were produced by laser ablation of a strontium titanate ($\rm{SrTiO_{3}}$) target \cite{leibrandt2007laser,osada2022deterministic} and photoionized by $\SI{461}{\nano m}$ and $\SI{405}{\nano m}$ light. The loading probability was controlled by adjusting the ablation laser intensity. All experiments reported here used  a single $^{88}\mathrm{Sr}^{+}$. Doppler cooling and fluorescence detection used the $S_{1/2}\leftrightarrow P_{1/2}$ transition at $\SI{422}{\nano m}$, together with a 1092 nm repumping laser on the  $D_{3/2}\leftrightarrow P_{1/2}$ transition. The qubit states were $\ket{\downarrow}=\ket{S_{1/2},m=-1/2}$ and $\ket{\uparrow}=\ket{D_{5/2},m=-5/2}$. Sideband cooling and coherent qubit control used the $\SI{674}{\nano m}$ electric-quadrupole transition~\cite{akerman2012quantum}, with a $\SI{1033}{\nano m}$ laser quenching population from $D_{5/2}$ to $S_{1/2}$. Permanent magnets outside the chamber provided a $\SI{2.58}{G}$ magnetic field perpendicular to the $\SI{674}{\nano m}$ laser wave vector.  The 674 nm laser was stabilized to an ultra-low-expansion (ULE) cavity. Long-term cavity drift was compensated by
comparison with a GPS-locked optical frequency comb. Approximately $\SI{10}{mW}$ of optical power at the ion yielded a carrier Rabi frequency $\Omega_{0}/2\pi$ of up to $\SI{320}{kHz}$. Continuous-wave sideband cooling \cite{reed2024comparison} prepared the
motion near its ground state, with $n_{\rm{th}}\cong0.19$ (Appendix C). The measured secular frequencies were
$(\omega_x,\omega_y, \omega_z)/2\pi\cong(1.13,1.30,0.35)\ \si{\mega Hz}$. The $x$ and $y$ modes were radial, and the $z$ mode was axial.
Compensation was performed along $x$, for which $\eta_{x}\cong0.056$.

\subsection{Calibration of the motional drive frequency}
Equation~(\ref{Equation3}) shows that the coherent-state amplitude depends not only on the ion displacement $x_{\rm d}$ but also on the detuning between the auxiliary drive and the secular motion. The auxiliary RF field can also shift the secular frequency (Appendix G), so the resonance frequency measured without the auxiliary drive does not necessarily coincide with that under the conditions used for micromotion compensation. We therefore calibrated the resonance frequency at the auxiliary-drive amplitude used for each compensation measurement. The calibration consisted of a broad frequency scan to approximately locate the resonance, followed by a motional echo measurement to refine its determination.

For the broad scan, the ion was prepared near its motional ground state and driven with a Gaussian electric-field pulse. Blue-sideband Rabi oscillations were then recorded at each drive frequency. Figure~\ref{figure3}(a) compares an off-resonant drive at $\omega_{\rm{d}}
/(2\pi) = \SI{1080}{kHz}$ with a near-resonant drive at $\SI{1126}{kHz}$. The near-resonant drive produced a faster Rabi oscillation, reflecting a larger coherent excitation at the same drive amplitude. The oscillations measured across the scan are shown in Fig.~\ref{figure3}(b). Sinusoidal fits yielded the effective Rabi frequencies plotted in Fig.~\ref{figure3}(c). Fitting their frequency dependence to the Gaussian response in Eq.~(\ref{Equation3}) gave a resonance frequency of $\omega_{x}/(2\pi) = (1126.8 \pm 0.2)\ \si{kHz}$. Because this procedure requires a Rabi-oscillation measurement at every frequency, it was used for the initial calibration. During compensation-voltage scans, the resonance was checked using the faster motional echo measurement.

\begin{figure}
\includegraphics[keepaspectratio, height=0.65\textwidth]{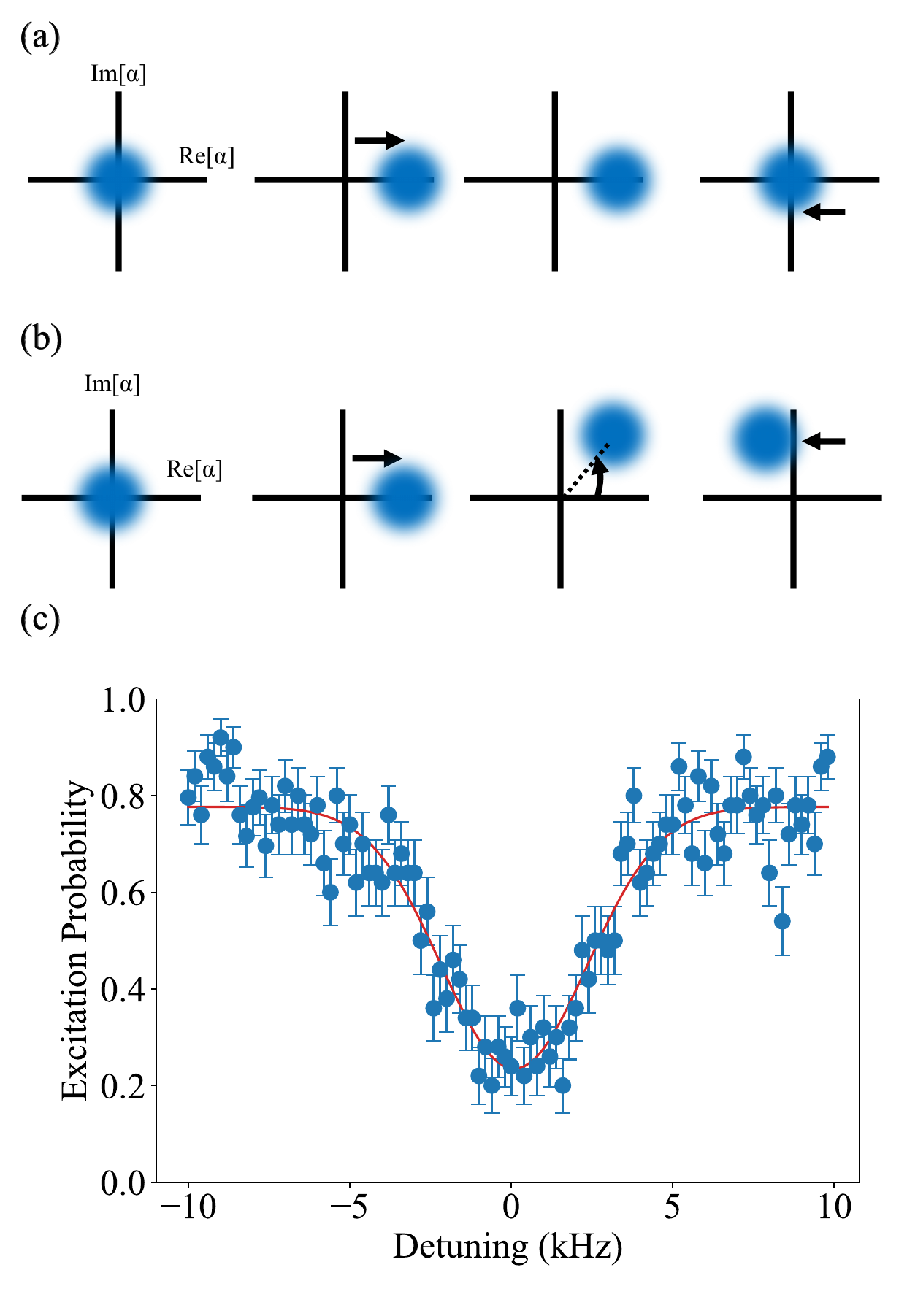}
\caption{\label{figure4}  (a) Phase space trajectory during the motional echo sequence for a resonant drive. The two phase-inverted drive pulses cancel each other, returning the motional state close to its initial state. (b) Phase-space trajectory for an off-resonant drive. Phase accumulation during the free-evolution interval results in an incomplete cancellation and a residual coherent displacement. (c) Measured spin-excitation probability versus detuning, with an empirical Gaussian-dip fit to the central minimum.
}
\end{figure}

The echo sequence is analogous to motional Ramsey interferometry \cite{mccormick2019coherently,wolf2019motional}. After cooling, two Gaussian electric-field pulses with opposite phases were applied with a delay $T = \SI{72}{\micro s}$. The Gaussian width of each pulse
was $\sigma_{\rm{d}}\cong\SI{5}{\micro s}$. A $\SI{5}{\micro s}$ blue-sideband probe pulse, approximately one-fifth of the $\pi$-pulse duration, was then
followed by qubit-state detection. At resonance, the two phase-space displacements cancel, returning the motion close to its initial state (Fig.~\ref{figure4}(a)). A detuned drive causes phase accumulation between the pulses and leaves a residual displacement (Fig.~\ref{figure4}(b)). In the weak-probe limit, the spin-excitation probability increases with the mean phonon number above its undriven baseline. Therefore, the excitation probability is minimized when the residual coherent-state amplitude vanishes. The central minimum in Fig. \ref{figure4}(c) was fitted with a Gaussian dip to determine the resonance frequency. This fit provides an empirical estimate of the minimum rather than a complete model of the echo spectrum.

\subsection{Micromotion compensation}
\begin{figure}
\includegraphics[keepaspectratio, height=0.7\textwidth]{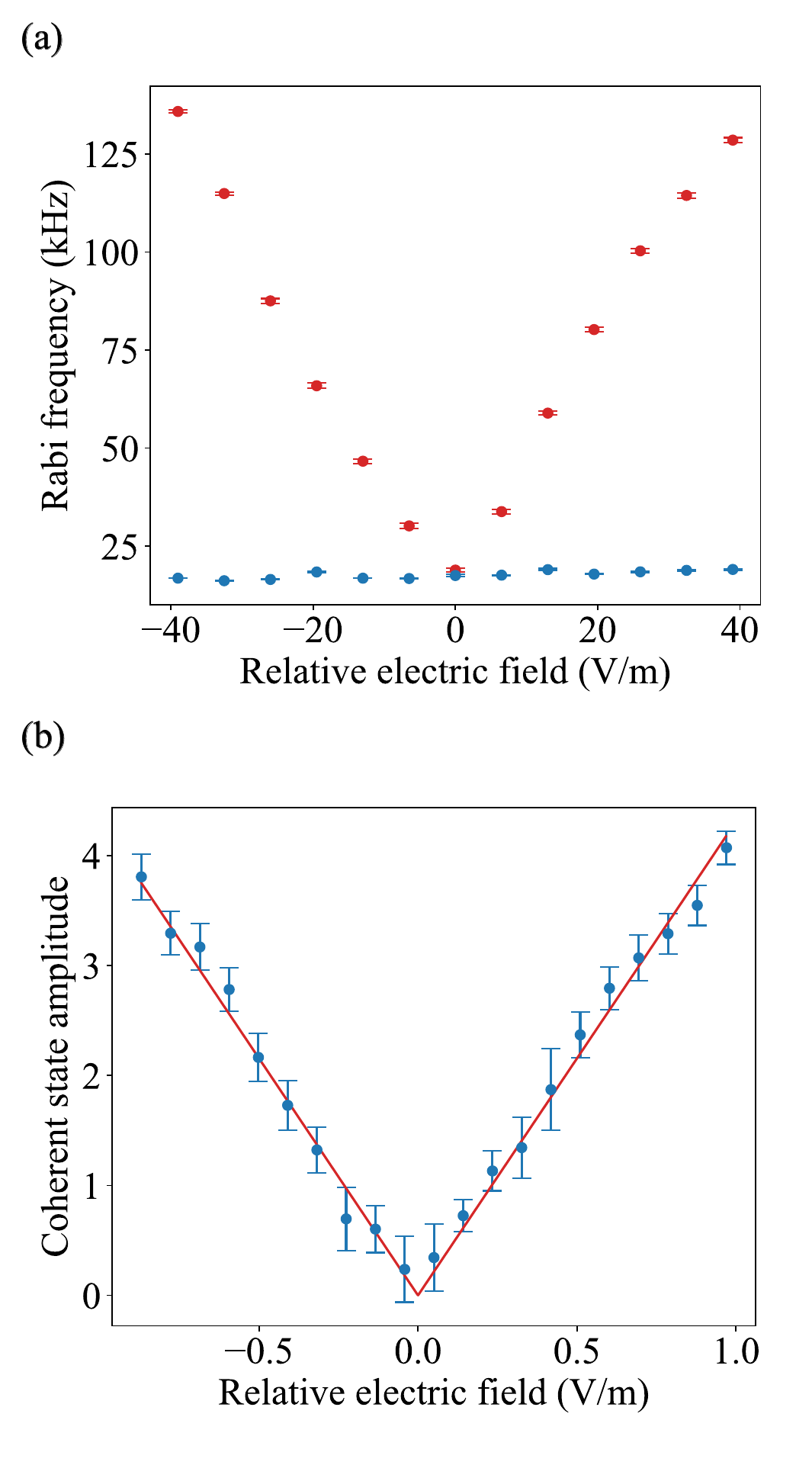}
\caption{\label{figure5}  (a) Micromotion compensation measurement in the regime of large displacement from the RF null. Red dots represent the estimated blue-sideband Rabi frequencies after electric drive and blue dots represent the estimated blue-sideband Rabi frequencies without electric drive. (b) Micromotion compensation measurement in the regime of small displacement from the RF null. The blue dots represent the estimated coherent-state amplitudes after the electric drive, and the red line represents a fit to Eq.~(\ref{Equation6}).
}
\end{figure}

We first measured the blue-sideband Rabi frequency with (red points) and without (blue points)  the auxiliary drive as a function of the compensation voltage (Fig. \ref{figure5}(a)). Electrode voltages were converted to electric fields with COMSOL Multiphysics\textsuperscript{\textregistered} \cite{COMSOL}. A voltage of 1 V was applied to the compensation electrodes, and the resulting field at the ion position defined the voltage-to-field conversion factor. The field origin in Fig. \ref{figure5}(a) was set at the compensation condition that minimized the driven Rabi frequency. Excess micromotion can change the carrier Rabi frequency through RF phase modulation of the ion-laser interaction. Because the sideband coupling depends on the carrier coupling, measuring the undriven blue-sideband Rabi frequency distinguishes changes in this coupling from excitation caused by the auxiliary drive. The undriven frequency remained nearly constant across the scan, whereas the driven frequency increased
away from the compensation minimum. This behavior is consistent with a displacement-dependent coherent excitation and indicates that variations in the underlying carrier coupling were small over the measured range.
For large coherent amplitudes, comparing the driven and undriven frequencies provided an approximate compensation signal.

Near the RF null, we extracted $|\alpha|$ by fitting the Rabi oscillations to the thermal-coherent-state model in Eq.~(\ref{Equation5}). A larger auxiliary-drive amplitude was used in this regime to increase the response to small displacements. Figure \ref{figure5}(b) shows the fitted amplitudes as a function of the compensation field. We fitted these data to
\begin{equation}
\label{Equation6}
|\alpha| = k|E-E_{\rm{opt}}|    
\end{equation}
where $k$ converts the field offset into the coherent-state amplitude, and $E_{\rm{opt}}$ is the field that minimizes the excitation. The fitted $E_{\rm{opt}}$ defines the field origin in Fig. \ref{figure5}(b). The voltage-to-field conversion was the same as in Fig. \ref{figure5}(a). The fit covariance matrix yielded a $1\sigma$ statistical uncertainty $\delta E_{\rm{opt}} = \SI{8}{mV/m}$. This is conditional on the fixed calibration parameters and the assumed model; uncertainties in those parameters and systematic effects were not propagated into this value. Using the harmonic confinement along $x$, the corresponding displacement uncertainty
is $\delta x_{\rm{d}} = \frac{|q| \delta E_{\mathrm{opt}}}{m\omega_{x}^{2}} = \SI{170}{\pico m}\cong0.02\times x_{\mathrm{zpf}}$. The measurement therefore determines the compensation minimum with subnanometer statistical precision. 

\section{Conclusion}

We demonstrated mode-selective micromotion compensation by mapping the ion displacement onto the coherent amplitude of a selected motional mode prepared near its ground state. An auxiliary drive applied to the RF electrodes generated a force proportional to the displacement, and qubit-assisted sideband measurements determined the resulting excitation. The optimal compensation field was obtained with a $1\sigma$ statistical uncertainty of $\SI{8}{mV/m}$, corresponding to approximately $\SI{170}{pm}$ in position. This uncertainty describes the precision of the fitted minimum and excludes systematic and calibration uncertainties. Near the minimum, small coherent excitations produce only weak changes in the blue-sideband dynamics, making finite qubit-readout statistics a limitation on the fitted displacement amplitude. Increasing the auxiliary-drive amplitude or duration increases the response to a given displacement. In the present setup, the useful duration was limited by motional coherence and the amplitude by the available RF amplifier gain. Both were chosen near their experimentally accessible limits.

Beyond these technical improvements, the use of nonclassical motional states provides another route to improve sensitivity. In the present experiment, the residual electric field was detected through a coherent motional state. Previous studies have shown that Fock states \cite{wolf2019motional}, squeezed states \cite{burd2019quantum}, and Schrödinger-cat states \cite{hempel2013entanglement} can enhance the sensitivity to small coherent-state amplitude. Because the present method converts the displacement into a coherent-state amplitude of the motional state, these quantum-enhanced sensing protocols can be naturally incorporated into the present scheme. Such an extension could improve the sensitivity beyond that achieved with the coherent-state measurement demonstrated here.

As the sensitivity is improved, however, a systematic limitation specific to the present technique becomes increasingly relevant. This limitation arises from a possible mismatch between the RF null of the auxiliary drive field and that of the trapping RF field. Such a mismatch can arise from nonuniform RF leakage onto the DC electrodes. Because the impedance of the leakage path to each DC electrode is frequency dependent, the leakage amplitudes can differ between the auxiliary and trapping RF frequencies. The resulting difference in the spatial electric-field distributions can lead to a relative displacement of the two RF nulls. 
For the trap used in this work, we estimate the RF-null mismatch to be approximately $61~\mathrm{pm}$ under the assumptions described in Appendix A, which is about three times smaller than the statistical uncertainty in the RF-null position achieved in the present experiment. This suggests that the mismatch is unlikely to be a dominant source of error at the current level of precision, although a more complete characterization of the leakage fields would be required to establish a rigorous systematic uncertainty.

\begin{acknowledgments}
We thank Atsushi Yamaguchi for insightful discussions on ULE-cavity stabilization and Lao Qifeng, JaEun Kim, and Yuta Matsumoto for fruitful discussions on ion trapping. This work has been supported by the JST Moonshot R$\&$D Program (Grant No. JPMJMS256G), JST ASPIRE (Grant No. JPMJAP2411), and JST SPRING (Grant No. JPMJSP2108).

\end{acknowledgments}

\section*{Data Availability}
The data supporting the findings of this study are available from the corresponding author upon reasonable request.

\appendix

\section{ESTIMATE OF THE RF-NULL MISMATCH}

\begin{figure}[b]
\includegraphics[keepaspectratio, height=0.35\textwidth]{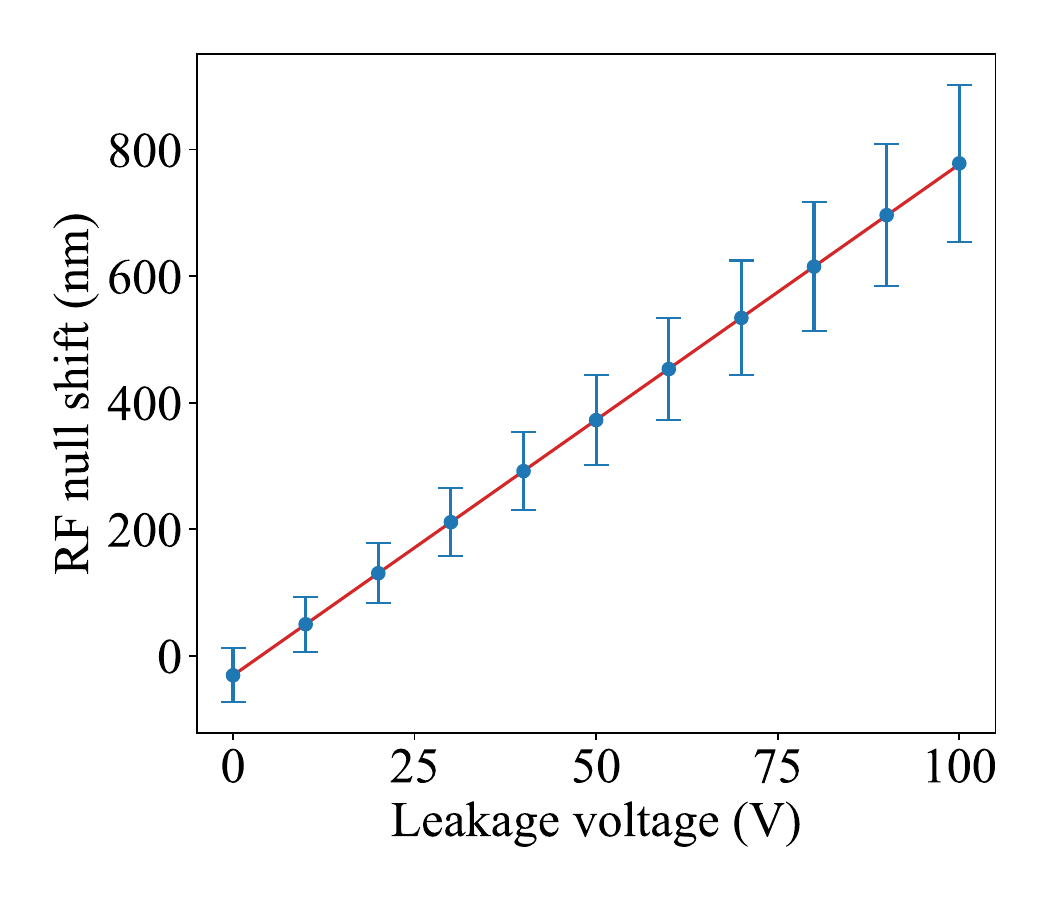}
\caption{\label{figure6} Simulation results of the RF null mismatch as a function of the leakage voltage amplitude. The blue circles represent numerical results obtained from finite-element simulations, and the red solid line represents a fit to these data. The error bars indicate the $1\sigma$ uncertainties obtained from the fits used to determine the RF-null positions from the simulated pseudopotentials.}
\end{figure}

A potential source of systematic error in the present micromotion compensation technique is a mismatch between the RF null associated with the auxiliary low-frequency drive and that of the trapping RF field. In the absence of RF leakage onto the DC electrodes, both fields are generated by the same RF electrodes and therefore share the same RF-null position. In practice, however, a finite amount of RF leakage remains. Because the impedance to each DC electrode is frequency dependent, the leakage amplitude can differ between the trapping RF and the auxiliary RF drive. These differences can modify the spatial electric-field distributions of the two RF fields and thereby shift their RF-null positions relative to each other. If the residual RF leakage were identical for all DC electrodes, the symmetry of the trap would be preserved and no relative shift of the RF nulls would occur. A mismatch arises when the RF leakage varies from electrode to electrode and breaks this symmetry. To estimate the resulting systematic error, we measured RF leakage through low-pass filters on all DC electrodes and combined these measurements with finite-element simulations performed by COMSOL Multiphysics\textsuperscript{\textregistered}~\cite{COMSOL}. We first calculated the RF-null displacement caused by nonuniform RF leakage. To estimate the sensitivity of the RF-null position to nonuniform leakage, we considered a simplified scenario in which leakage was applied to the DC electrode producing the largest electric field at the ion position. Figure~\ref{figure6} shows the calculated RF-null displacement as a function of the leakage-voltage amplitude. The trapping RF amplitude was fixed at $\SI{200}{V}$, while the leakage amplitude was varied from 0 to $\SI{100}{V}$. The blue circles show the results of the finite-element simulations. For each leakage amplitude, the calculated pseudopotential was fitted with a polynomial function, from which the RF-null position was determined. The error bars indicate the uncertainty of this fit. The red line shows a linear fit to the simulated RF-null displacement as a function of the leakage amplitude, yielding a conversion coefficient of $\SI{8.078}{\pico m/\milli V}$ with a $1\sigma$ fitting uncertainty of $\SI{0.004}{\pico m/\milli V}$.
The measured RF leakage after the low-pass filters had a mean amplitude of $\SI{150}{\milli V}$ and a standard deviation of $\SI{7.5}{\milli V}$ across the DC electrodes. Using the measured standard deviation of the leakage amplitudes across the DC electrodes ($\SI{7.5}{\milli V}$) and the simulated conversion coefficient ($\SI{8.078}{\pico m/\milli V}$), we estimate the relative displacement of the two RF nulls to be $\SI{7.5}{mV}\times\SI{8.078}{\pico m/\milli V}\simeq\SI{61}{pm}$. This estimate assumes that the electrode-to-electrode variation in leakage amplitude is the dominant source of the mismatch and does not explicitly account for the relative phases or combined spatial contributions of the leakage fields. The estimated mismatch is approximately three times smaller than the statistical uncertainty in the RF-null position obtained in the present experiment, suggesting that this effect is unlikely to be the dominant limitation at the current level of precision.

\begin{figure}
\includegraphics[keepaspectratio, width=0.48\textwidth]{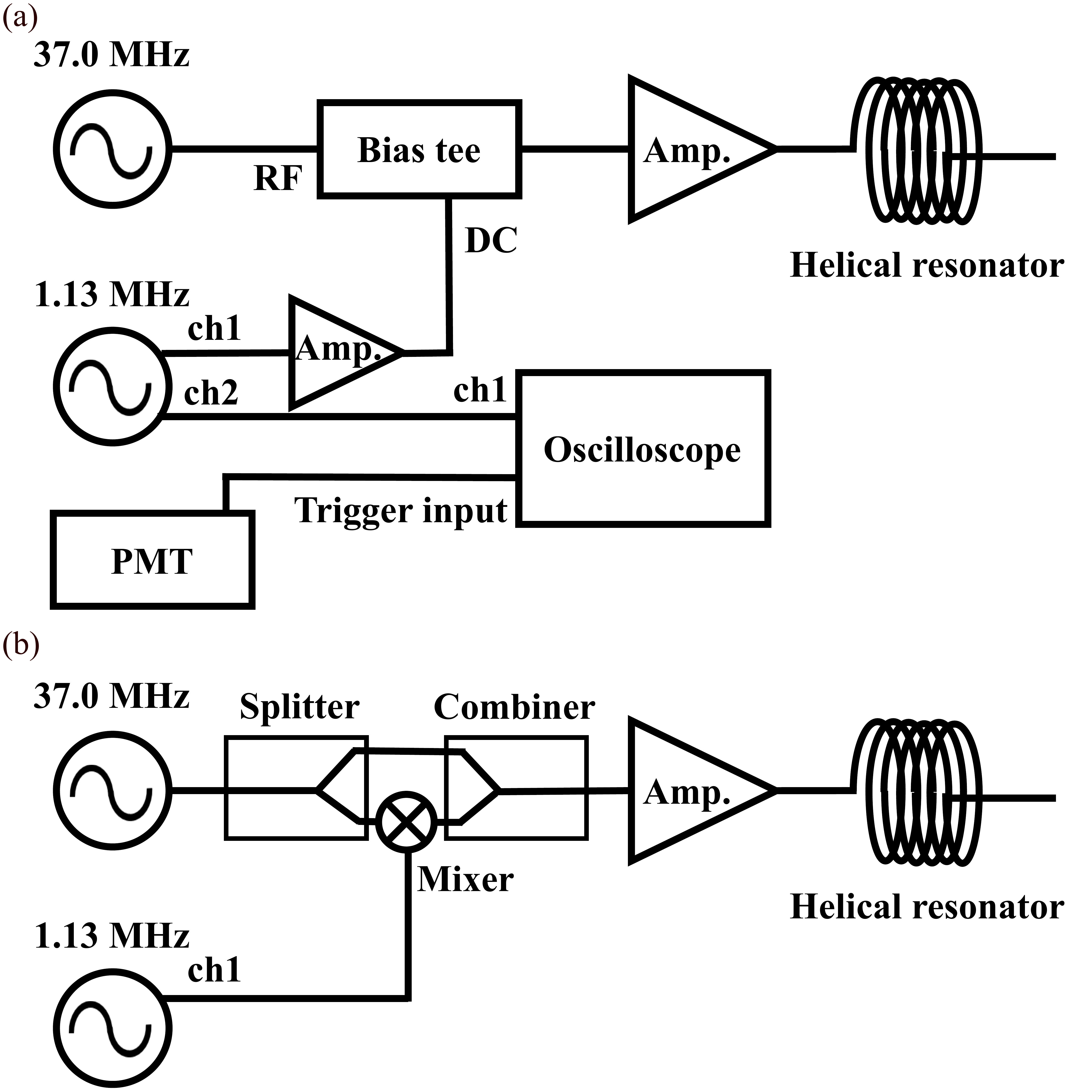}
\caption{\label{figure7} (a) Circuit diagram for trap RF and coherent drive. Amp. represents Amplifier. (b) Circuit diagram for amplitude modulation of the trap RF signal. This circuit diagram was only used in Appendix F.}
\end{figure}

\section{ELECTRICAL CONFIGURATION}

Figure~\ref{figure7}(a) shows the circuit configuration used in this work. The trapping RF signal was generated by a Rohde $\&$ Schwarz SMB100A signal generator, while the auxiliary low-frequency RF signal used to drive the secular motion of the ion was produced by channel 1 of a Zurich Instruments HDAWG. The auxiliary RF signal was amplified by a Mini-Circuits ZHL-500LN+ amplifier and combined with the trapping RF signal by a bias tee. The combined signal was then amplified by a Mini-Circuits ZHL-20W-13+ amplifier and applied to the trap through a helical resonator. The resonator had a resonance frequency of $\SI{37.0}{\mega Hz}$ and a loaded quality factor of $\rm{Q}=140$. For the correlation measurements described in Appendix E, the motional drive source was replaced by an NF WF1974 waveform generator. As in the main experiment, channel 1 was combined with the trapping RF through a bias tee and applied to the RF electrodes. Channel 2 generated a signal with the same frequency and phase as channel 1 and was connected to a Rohde $\&$ Schwarz RTM3004 oscilloscope. The oscilloscope was used to measure the correlation between the ion fluorescence and the RF signal driving the secular motion. The fluorescence photons detected by the photomultiplier tube (PMT) were used as the trigger signal, while the reference RF signal was recorded and averaged over many repetitions. If the electrical pulses from the PMT and the applied RF drive are uncorrelated, the averaged waveform approaches zero. In contrast, when the fluorescence exhibits a modulation synchronized with the applied RF drive, a finite oscillatory signal remains after averaging. The amplitude of this residual signal therefore provides a measure of the correlation between the fluorescence and the motional drive field.

Figure~\ref{figure7}(b) shows the circuit diagram used for the measurements described in Appendix F. The RF signal generated by a Rohde $\&$ Schwarz SMB100A signal generator was split into two paths by a Mini-Circuits ZSC-2-1+. One path was fed into a Mini-Circuits ZAD-3+ mixer together with the output of the HDAWG. The modulated RF signal from the mixer was then recombined with the unmodulated RF signal from the other path by a ZSC-2-1+. The resulting signal was applied to the RF electrodes through the helical resonator. This configuration allowed us to apply amplitude modulation to the trapping RF signal.

\begin{figure}
\includegraphics[keepaspectratio, width=0.45\textwidth]{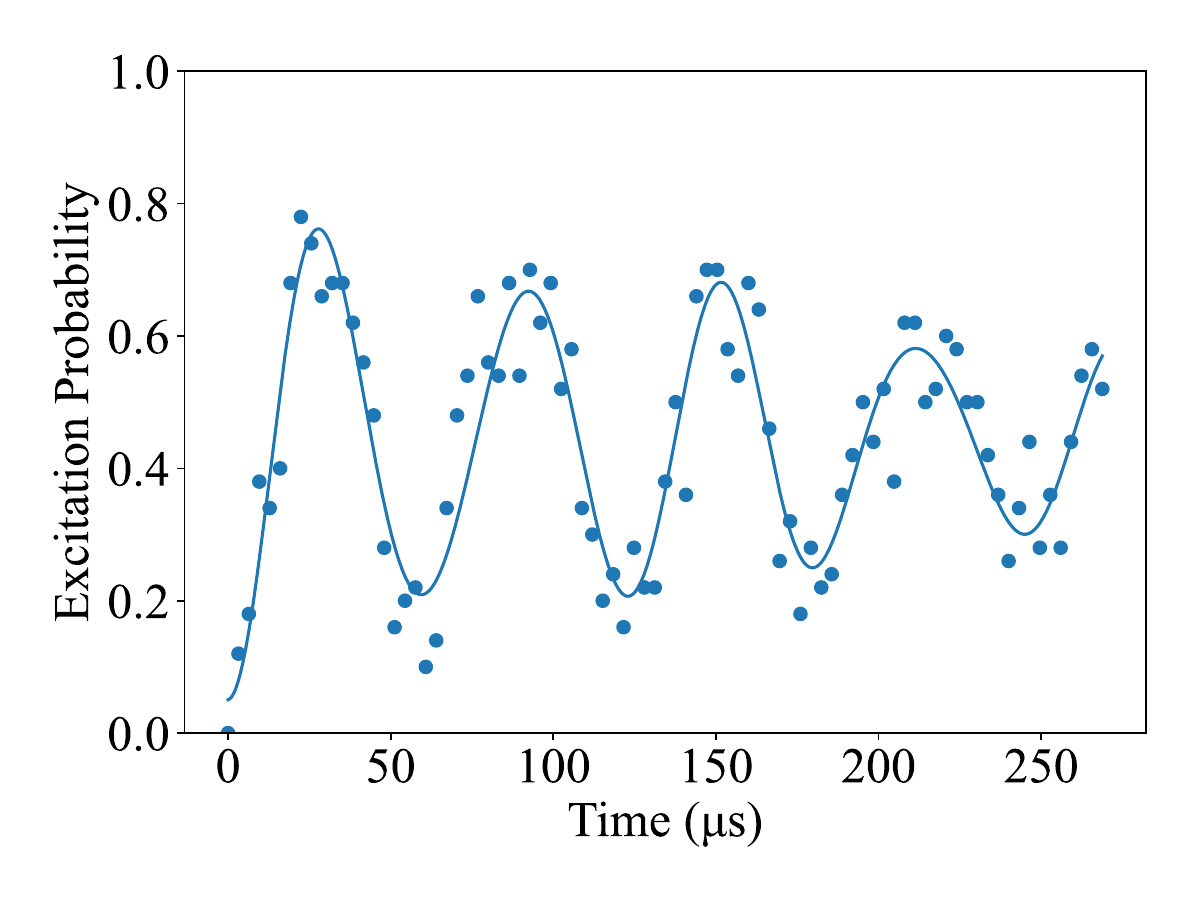}
\caption{\label{figure8} Blue-sideband Rabi oscillations after sideband cooling (blue circles) and the corresponding fit (solid line).}
\end{figure}

\section{CALIBRATION OF THE BLUE-SIDEBAND RABI OSCILLATIONS}
To estimate the coherent-state amplitude from the blue-sideband Rabi oscillations, the spin-initialization probability, the initial mean phonon number, and the blue-sideband Rabi frequency were determined beforehand. The dynamics of the blue-sideband Rabi oscillation were modeled as
\begin{equation}
    \label{equation:C1}
    {P}_{\uparrow}(t) = \frac{p_{\rm{init}}}{2}\sum_{n=0}^{N}
    p_{n}(n_{\rm{th}})\big\lbrack 1-e^{-\gamma t}\cos(\eta_{x}\Omega\sqrt{n+1}t) \big\rbrack
\end{equation}
where \(p_n(n_{\rm{th}})\) is the phonon-number distribution. After sideband cooling, the motional state was assumed to follow a thermal distribution,
\begin{equation}
\label{equation:C2}
    p_{n}(n_{\rm{th}})=\frac{n_{\rm{th}}^n}{(n_{\rm{th}}+1)^{n+1}}.
\end{equation}
Figure~\ref{figure8} shows the measured blue-sideband Rabi oscillations together with the fit to Eqs.~(\ref{equation:C1}) and (\ref{equation:C2}). From the fit, we obtained $p_{\rm init}=0.79\pm0.07$, $n_{\rm th}=0.19\pm0.07$, and $\eta_x\Omega_0/2\pi=16.45\pm\SI{0.07}{kHz}$. In the analysis of Fig.~\ref{figure5}(b), these parameters were fixed, and the phonon-number distribution $p_n$ in Eq.~(\ref{equation:C1}) was replaced by that of a displaced thermal state. The coherent-state amplitude $|\alpha|$ was then obtained as the fitting parameter from the blue-sideband Rabi oscillations measured after the motional drive.


\begin{figure}
\includegraphics[keepaspectratio, height=0.7\textwidth]{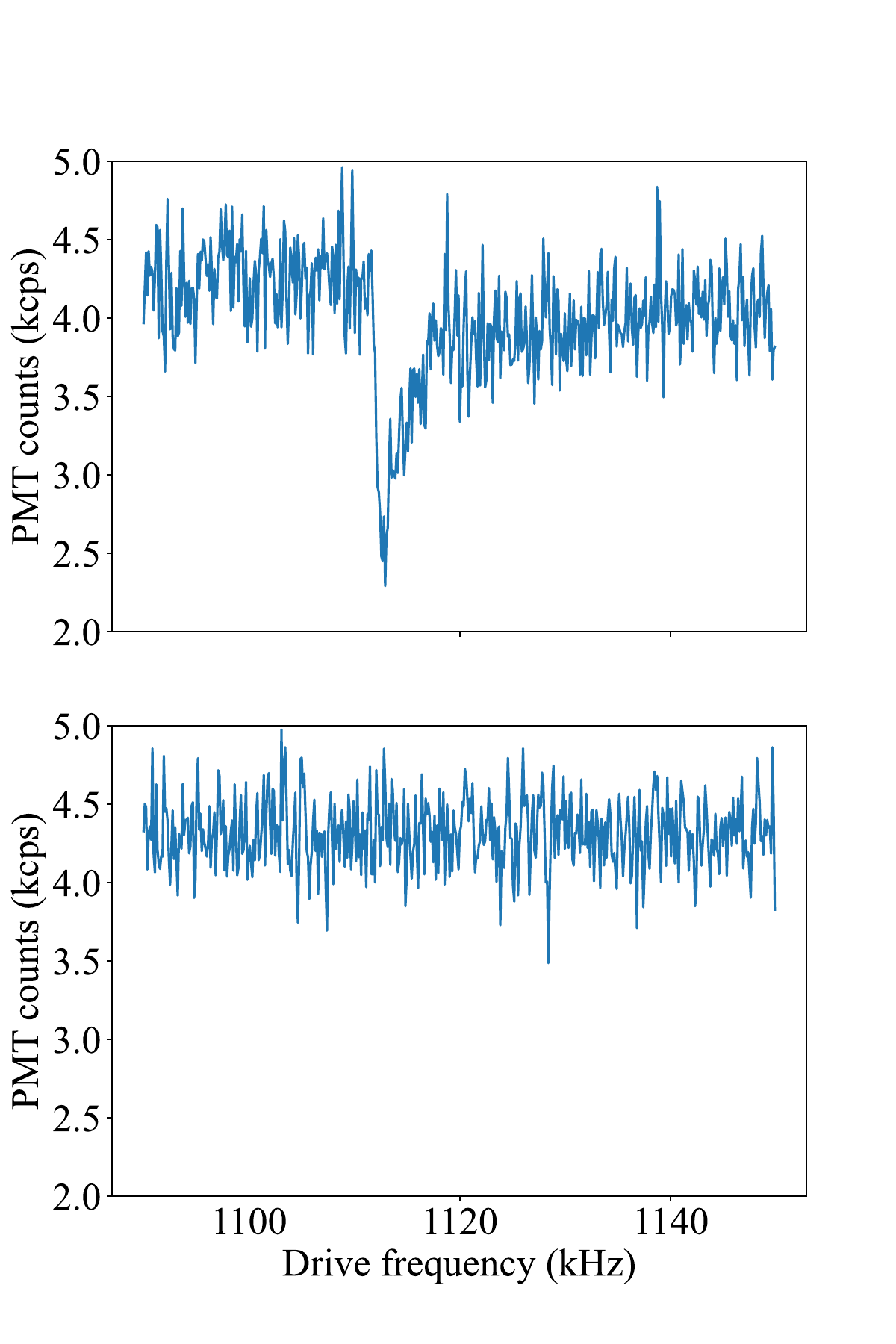}
\caption{\label{figure9} (a) Tickle spectrum measured with the ion displaced from the RF null. (b) Tickle spectrum measured near the optimal compensation condition. The secular frequencies differ from those in Fig. \ref{figure3} because the data were acquired on a different day.}
\end{figure}

\section{TICKLE MEASUREMENT}
In the present work, stray DC electric fields are detected by a coherent motional excitation. Maximizing the sensitivity of this approach requires accurate calibration of the secular frequency, which was performed by the motional echo technique described in the main text. One of the methods for measuring the secular frequency of trapped ions is the tickle method~\cite{yu1993noise}. 

Figure~\ref{figure9}(a) shows a secular frequency measurement obtained by the tickle method in the presence of a stray DC electric field of approximately $40~\mathrm{V/m}$. As evident from the spectrum, the resonance lineshape is significantly distorted. This distortion originates from the anharmonicity of the trapping potential, which becomes relevant when the ion is driven to large oscillation amplitudes. One advantage of the tickle method is that it is insensitive to frequency fluctuations of both the qubit transition and the 674 nm laser. However, the method relies on detecting changes in the ion fluorescence and therefore requires a relatively large motional excitation. As a result, the measured spectrum can be strongly affected by trap anharmonicities, as observed in Fig.~\ref{figure9}(a). For this reason, the secular frequency in the present work was calibrated by the motional echo sequence rather than the tickle method.

Nevertheless, the tickle method can still provide a qualitative indication of the micromotion compensation condition. Figure~\ref{figure9}(b) shows a tickle spectrum measured with the DC compensation voltage set near its optimal value. In contrast to Fig.~\ref{figure9}(a), no discernible resonance dip is observed. This result is consistent with a substantial reduction of the ion displacement. Consequently, although the absence of a tickle resonance serves as a useful qualitative indicator of micromotion compensation, it cannot be used to determine the residual electric field with the precision achieved by the coherent-state-based method presented in this work.

\begin{figure}
\includegraphics[keepaspectratio, height=0.35\textwidth]{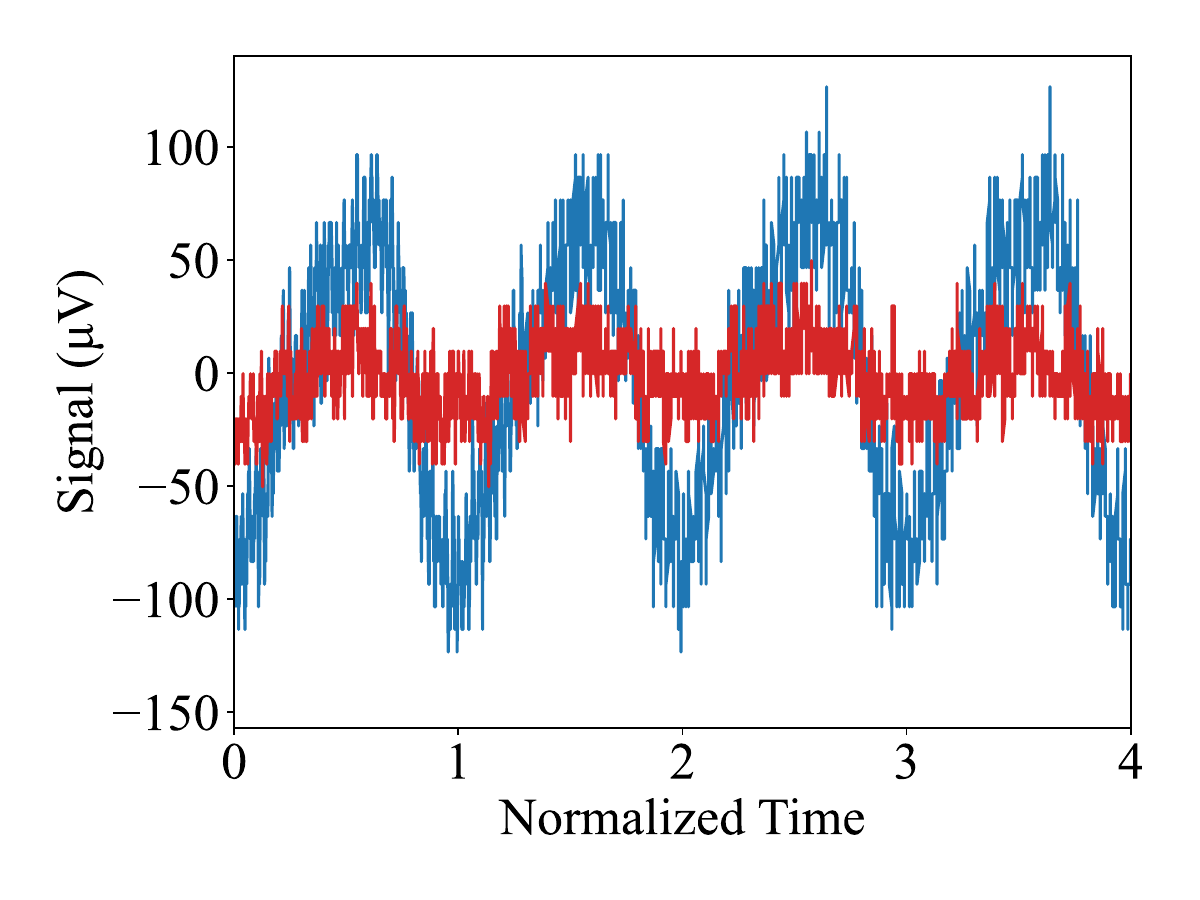}
\caption{\label{figure10} Correlation signals near the optimal compensation field (red solid line) and with an additional stray field of approximately 130 V/m applied (blue solid line).}
\end{figure}

\section{FLUORESCENCE CORRELATION WITH DRIVEN SECULAR MOTION}
One of the most widely used techniques for micromotion compensation is the photon-correlation method. In this approach, the magnitude of the ion micromotion is inferred from the correlation between the fluorescence signal emitted by the ion and the RF signal used for trapping.

In the method presented in this work, the ion displacement is converted into a motional excitation through an oscillating electric field whose effective driving strength is proportional to the displacement. Motivated by the similarity to the conventional photon-correlation technique, we investigated an alternative implementation based on continuous-wave motional driving. Using the setup shown in Fig.~\ref{figure7}(a), the oscillating electric field was applied continuously, and the correlation between the applied drive and the fluorescence signal was measured to estimate the ion displacement. Figure~\ref{figure10} shows the measured correlation signals near the optimal compensation field (red solid line) and with an additional stray field of approximately 130 V/m applied to intentionally displace the ion from the RF null (blue solid line). The horizontal axis represents time normalized by the drive frequency. The measurements were performed with a drive amplitude corresponding to $\SI{20}{V}$. As evident from the comparison, the correlation signal increases when the ion is displaced from the RF null. This behavior arises because the effective motional drive is proportional to the ion displacement, leading to enhanced modulation of the fluorescence signal. However, the signal-to-noise ratio was significantly lower than that of the conventional photon-correlation technique. We attribute this reduction in sensitivity to the lower frequency of the secular motion compared to the trap RF drive. Photon-correlation measurements are based on fluorescence modulation induced by the Doppler effect associated with the ion velocity. Consequently, the signal amplitude is determined by the velocity modulation rather than the displacement amplitude itself. Because the secular frequency is much smaller than the trap RF frequency, a driven secular motion produces a smaller velocity modulation for a given displacement amplitude. The resulting Doppler modulation is therefore weaker, leading to a reduced signal-to-noise ratio compared with conventional photon-correlation measurements.

\begin{figure}
\includegraphics[keepaspectratio, height=0.35\textwidth]{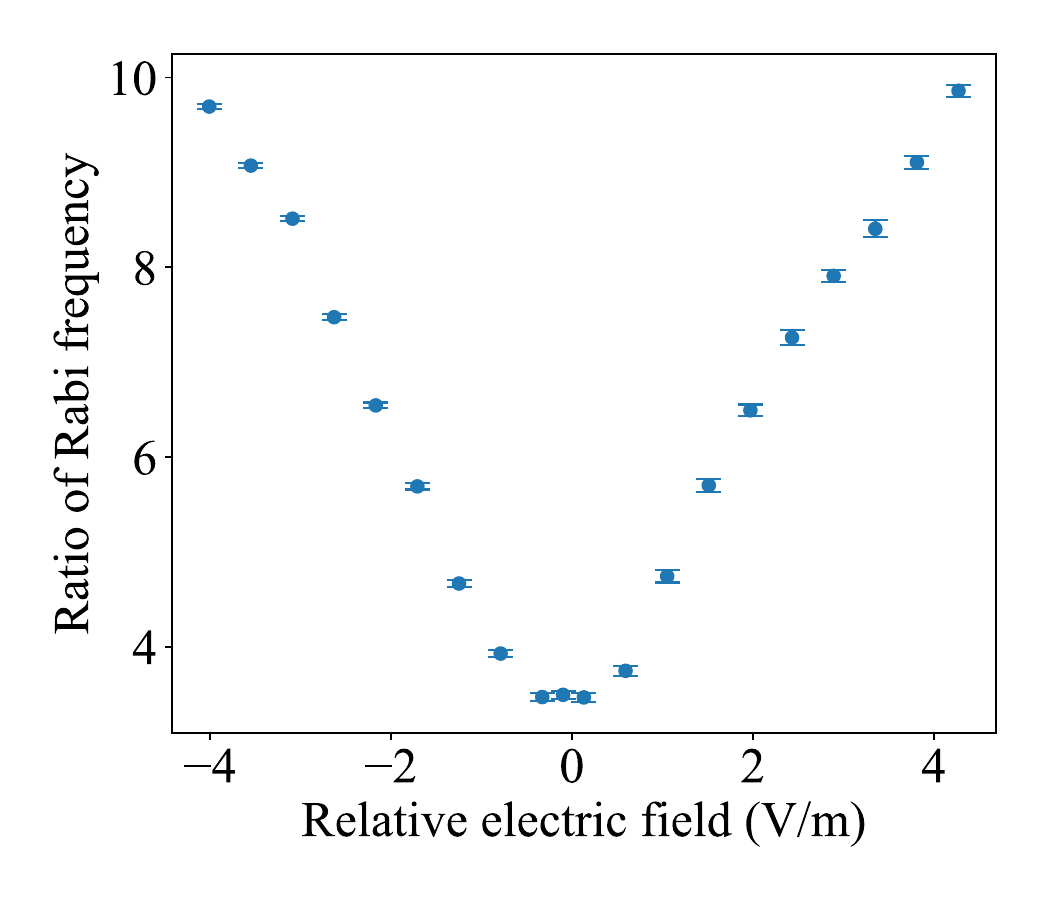}
\caption{\label{figure11} Blue dots represent the ratio of the blue-sideband Rabi frequency measured after the amplitude modulation pulse to that measured without amplitude modulation.}
\end{figure}

\section{MOTIONAL EXCITATION BY TRAPPING-RF AMPLITUDE MODULATION}
In the method employed in this work, an auxiliary RF signal resonant with the secular frequency is applied to the RF electrodes to generate a coherent motional excitation whose amplitude is proportional to the ion displacement. An alternative approach to generating such an excitation is to modulate the amplitude of trapping RF at the secular frequency \cite{nadlinger2021micromotion}. Figure~\ref{figure11} shows the micromotion compensation measurement obtained by this amplitude modulation scheme with the circuit shown in Fig.~\ref{figure7}(b). The horizontal axis represents the relative compensation electric field applied to the ion, while the vertical axis shows the ratio of the blue-sideband Rabi frequency measured after the amplitude modulation pulse to that measured without amplitude modulation. As shown in Fig.~\ref{figure11}, the Rabi-frequency ratio does not reach unity even near the optimal compensation field, indicating a finite coherent excitation. This observation shows that a background coherent excitation remains under amplitude modulation even when the ion is located near the RF null. We further confirmed experimentally that the magnitude of this background excitation increases with the modulation index. The physical origin of this excitation has not yet been identified. Such background excitation produces a finite signal even at the RF null and can therefore reduce the accuracy with which the optimal compensation field is determined. For this reason, we employed the auxiliary-RF scheme described in the main text, in which we observed no detectable background excitation.


\section{SECULAR-FREQUENCY SHIFT CAUSED BY THE AUXILIARY DRIVE}

\begin{figure}
\includegraphics[keepaspectratio, height=0.35\textwidth]{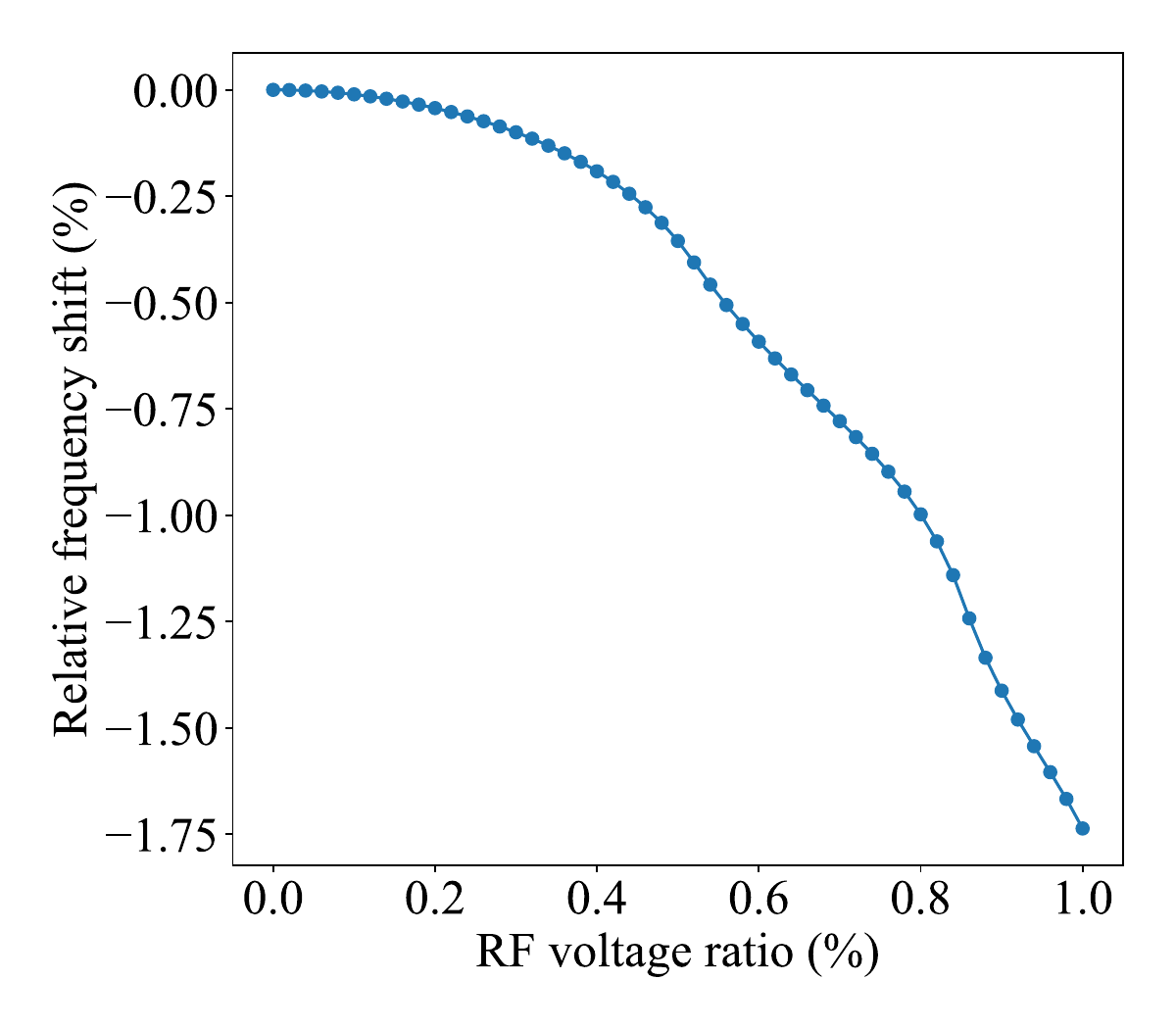}
\caption{\label{figure12} Relative secular-frequency shift induced by the two-frequency Paul-trap effect as a function of the amplitude ratio $V_{\rm d}/V_{\rm t}$ between the auxiliary and trapping RF fields. The blue circles show the numerical results, and the solid line connects the data points as a guide to the eye.}
\end{figure}

In the present method, an auxiliary RF signal is applied to the RF electrodes to resonantly drive the secular motion of the ion. The ion is therefore simultaneously subjected to two RF fields, and its motion can be described in terms of a two-frequency Paul trap \cite{trypogeorgos2016cotrapping}. As a consequence, the secular frequency in the presence of the auxiliary RF drive can differ from that obtained with the trapping RF alone. To evaluate this frequency shift, we numerically solved the classical equation of motion of the ion in the presence of both RF fields. The trap parameters were first chosen to reproduce the secular frequency measured experimentally for $V_{\rm d}=0$. We then increased the amplitude of the auxiliary RF drive and determined the secular frequency from the resulting ion motion. Figure~\ref{figure12} shows the secular-frequency shift calculated as a function of the amplitude ratio $V_{\rm d}/V_{\rm t}$ between the auxiliary and trapping RF fields. The secular frequency deviates from its value in the absence of the auxiliary RF drive as $V_{\rm d}/V_{\rm t}$ increases, owing to the two-frequency Paul trap contribution.
Consequently, the secular frequency measured in the absence of the auxiliary RF drive does not necessarily coincide with the resonance frequency under the conditions used for micromotion compensation. Because the coherent-state amplitude is sensitive to the detuning between the auxiliary RF drive and the secular motion, neglecting this frequency shift can reduce the sensitivity to the ion displacement. We therefore experimentally calibrated the resonance frequency using the same auxiliary RF amplitude as that employed in the micromotion compensation measurements, rather than using the secular frequency measured without the auxiliary RF drive. The frequency-calibration procedure described in the main text thus determines the actual resonance condition while accounting for the frequency shift associated with the two-frequency Paul trap.

\nocite{*}

\bibliography{apssamp}

\end{document}